\documentclass[12pt]{iopart}
\usepackage{graphicx}
\begin{document}

\title[Mott-Hubbard transition]{Mott-Hubbard 
transition of cold atoms in optical lattices}

\author{Wilhelm Zwerger}

\address{ Sektion Physik, Universit\"at M\"unchen 
Theresienstr. 37
D-80333 M\"unchen, Germany}

\begin{abstract}
We discuss the superfluid to Mott-insulator transition 
of cold atoms in optical lattices recently observed by
Greiner et.al. (Nature 415, 39 (2002)). The 
fundamental properties of both phases and their 
experimental signatures are discussed
carefully, including the limitations of
the standard Gutzwiller-approximation.
It is shown that in a one-dimensional
dilute Bose-gas with a strong transverse confinement (Tonks-gas),
even an arbitrary weak optical lattice is able to induce a Mott like
state with crystalline order, provided the dimensionless interaction 
parameter is larger than a critical value of order one. The 
superfluid-insulator transition of the Bose-Hubbard model in this case 
continuously evolves into a transition of the commensurate-
incommensurate type with decreasing strength of the external 
optical lattice.

\end{abstract}


\maketitle

\section{Introduction}

The realization of BEC in ultracold atomic gases [1-3] has opened a wide 
area of research in atomic physics, where quantum-statistical effects
are of crucial importance: upon cooling, bosonic atoms in a trap
condense into a superfluid state at a rather sharply defined critical 
temperature $T_c$ while fermionic atoms continuously evolve into a
degenerate noninteracting gas, resisting spatial compression due to their
Fermi pressure [4]. Both features are a consequence of the purely 
statistical interaction between the atoms. By contrast, the actual 
interparticle potential plays only a comparatively minor role. To see
this, we start from the standard pseudopotential description, which 
replaces the complicated interatomic potential by an effective contact 
interaction of the form 

$$U(\vec x)=\frac{4\pi\hbar^2 a_s}{m}\cdot\delta(\vec x)
=g\cdot\delta(\vec x)
\eqno(1)$$

containing the exact s-wave scattering length $a_s$ as the only parameter. 
For identical fermions, there is no s-wave scattering due to the
Pauli-principle and thus we obtain an ideal Fermi gas to lowest order. 
For bosons, in turn, $a_s$ is finite, however at a given density $n$
the importance of direct interaction effects can be estimated from the 
ratio

$$\gamma=\frac{\epsilon_{int}}{\epsilon_{kin}}=
\frac{gn}{\hbar^2 n^{2/3}/m}\approx n^{1/3}\, a_s\eqno(2)$$

between the interaction and the kinetic energy per particle. Now the average
interparticle spacing $n^{-1/3}$ is usually much larger than the 
scattering length and thus $\gamma$ is very small, with typical values
around $0.02$. This puts us into the weak coupling limit where the many body
ground state of $N$ bosons is well approximated by a simple product [5]

$$\Psi_{GP}\left( \vec x_1, \vec x_2, \ldots \vec x_N\right)\,=\,\prod_{i=1}^N\,
\phi(\vec x_i)\eqno(3)$$

in which all atoms are in the identical single particle state 
$\phi(\vec x)$. Taking (3) as a variational ansatz, the optimal 'macroscopic
wave function' $\phi(\vec x)$ is found to obey the well known Gross-Pitaevski
equation. It describes - even on a quantitative level - a wealth of
remarkable and nontrivial properties of trapped condensates from
interference between different condensates [6] to collective modes [7]
or vortices [8]. From a many body point of view, the effective single 
particle or Hartree-description (3) is of course the simplest of all 
possible cases, containing no interaction induced correlations between
different atoms at all. A first step to go beyond this mean field 
description is the well known Bogoliubov theory. This is usually
introduced by considering small fluctuations around the Gross-Pitaevski
equation in a systematic expansion in the number of noncondensed
particles [9]. As emphasized, for example, by Leggett [5], it is more instructive 
from a many body point of view to formulate Bogoliubov theory
in such a way that the many body boson ground state is
approximated by an optimized product 

$$\Psi_{Bog.}\left( \vec x_1, \vec x_2, \ldots \vec x_N\right)\,=\,\prod_{i<j}
\phi_2(\vec x_i,\vec x_j)\eqno(4)$$

of identical, symmetric two particle wave functions $\phi_2$. This allows us
to build in the interaction beyond the mean field potential by suppressing 
configurations in which particles $i$ and $j$ are close together. The 
many body state thus incorporates two-particle correlations which are 
important e.g. to obtain the standard sound modes and the related
coherent superposition of 'particle' and 'hole' excitations [10]. However,
even the Bogoliubov description is restricted to the regime $\gamma\ll 1$,
where interactions lead only to a small depletion of the condensate 
at zero temperature. The associated ground state is again characterized by 
a macroscopic matter wave field and continuously evolves from that of a
noninteracting gas.

An obvious way to go beyond the weak coupling regime is to increase
the dimensionless interaction strength parameter $\gamma$ by increasing
the scattering length via a Feshbach resonance. Of course, this method
is limited by the fact that the associated condensate lifetime strongly
decreases due to three-body losses which occur at a rate [11]

$$\dot n/n\, =-\hbox{const.}\frac{\hbar}{m}\left( na_s^2\right)^2\eqno(5)$$

In spite of this problem, this method of reaching the strong coupling 
regime has been followed quite successfully recently, see e.g. [12] 
and [13]. In the following, we will discuss
an alternative route to reach strong coupling even at small values
of $n^{1/3}a_s$. It is based on confining cold atoms in the periodic 
potential of an optical lattice generated via the dipole force
which atoms experience in a standing, off-resonant light field. Depending on 
the sign of the polarizability, the atoms are 
attracted either to the nodes or the anti-nodes of the laser intensity. In
this manner, one-, two-, or three-dimensional lattices can be created with 
a lattice constant $a=\lambda/2$ which is half the laser wavelength
(typically $a$ is in the range between $0.5\mu$m and $5\mu$m). In the 
simplest case, three orthogonal, independent standing laser fields
with wave vector $k$ produce a separable 3d lattice potential

$$V(x,y,z)=V_{0}\left(\sin^2{kx}+\sin^2{ky}
+\sin^2{kz}\right)\eqno(6)$$ 

with a tunable amplitude $V_0$. A convenient measure for the strength $V_0$
of the lattice potential is the recoil energy $E_r=\hbar^2k^2/2m$ which is
typically in the few kHz range. In a deep optical lattice with $V_0\gg E_r$,
the energy $\hbar\omega_0=2E_r\left(V_0/E_r\right)^{1/2}$ of
local oscillations in the well is much larger than the recoil energy 
and each well supports many quasi-bound states. For instance in the 
deepest lattices generated in the recent experiments by Greiner et.al 
[14] with $V_0=22E_r$ this number is around four while the local
oscillation frequency has reached $30$kHz.  Provided all the 
atoms are in the lowest vibrational level at each site, their motion is 
frozen except for the small tunneling amplitude to neighbouring sites. The
atoms are then effectively confined to move in the lowest band of the 
lattice. With $|\vec l>$ as the states localized at site $\vec l$, 
the appropriate single particle eigenstates are Bloch-waves 
$|\vec q>=\sum_l\,\exp{i\vec q\cdot\vec l}\; |\vec l>$ with 
quasimomentum $\vec q$ and energy

$$\epsilon(\vec q)=\frac{3}{2}\hbar\omega_0 -2J\left(\cos{q_xa}+\cos{q_ya}
+\cos{q_za}\right)\eqno(7)$$

The bandwidth parameter $J$ is essentially the gain in kinetic energy
due to nearest neighbour tunneling. In the limit $V_0\gg E_r$ it can be 
obtained from the exact result for the width of the lowest band in the 
1d Mathieu-equation 

$$J=\frac{4}{\sqrt\pi}E_r\left(\frac{V_0}{E_r}\right)^{3/4}
\exp{-2\left(\frac{V_0}{E_r}\right)^{1/2}}\eqno(8)$$

Obviously, in a lattice, it is $J$ which plays the role of the kinetic
energy per particle in the homogeneous case. The effective value of 
$\gamma=\epsilon_{int}/\epsilon_{kin}$ is therefore very large in 
optical lattices, increasing exponentially with $V_0/E_r$. Thus it is the 
quenching of the kinetic energy for motion in the lowest band which
drives cold atoms into the strong coupling regime, even though
$n^{1/3}a_s$ may still be much smaller than one. Alternatively, one may
argue that $\epsilon_{kin}$ becomes small because atoms in a deep
optical lattice have an exponentially large effective mass $m^*=\hbar^2/
2a^2J$. To really obtain interesting many body effects it is of course 
necessary to have the interaction and the kinetic energy of the same
order. This requires optical lattices in which the number of atoms per site
is of order one or larger, a regime which has been possible to reach only 
recently with Bose-Einstein-condensates.

\section{The superfluid- to Mott-insulator transition}

In the following, I will discuss
the Mott-Hubbard transition for bosonic atoms,
as a generic example
illustrating how cold atoms in optical lattices can be used to study 
genuine many body phenomena in dilute gases.
The original idea suggesting the possibility of nontrivial many body 
states using cold atoms in optical lattices is due to Jaksch et.al. [15].
Their starting point is the so called Bose-Hubbard model, originally
introduced by Fisher et.al. in a rather different context [16]. It 
describes bosons hopping with amplitude $J$ to nearest neighbors on
a regular lattice of sites $l$. The particles interact with a zero-range,
on-site repulsion $U$, disfavouring configurations with more than one atom 
at a given site. With $\hat b_l^{\dagger}$ as the creation operator of a 
boson at site $l$ and $\hat n_l=\hat b_l^{\dagger}\hat b_l$ the associated
number operator, the Hamiltonian reads

$$\hat H=-J\sum_{<ll'>}\,\hat b_{l}^{\dagger}\hat b_{l'}\, +
\,\frac{U}{2}\sum_{l}\,\hat n_{l}(\hat n_{l}-1)\, +\,
\sum_{l}\,\epsilon_{l}\hat n_{l}\eqno(9)$$ 

Here $<ll'>$ denotes a sum over nearest neighbour pairs, including 
double counting. The last term with a variable on-site energy 
$\epsilon_l$ is introduced to describe the effect of the trapping potential
and acts like a spatially varying chemical potential. The form of the 
interaction term is precisely that obtained by viewing  each site as
a local condensate with a Gross-Pitaevski mean field potential. The
relevant interaction parameter $U$ is thus given by an integral over the
on-site wave function $w(\vec x)$ via

$$U=g\int\,\vert w(\vec x)\vert^4 =\sqrt{\frac{8}{\pi}} ka_{s}E_{r}
\left(\frac{V_{0}}{E_{r}}\right)^{3/4}\eqno(10)$$

The explicit result is obtained by taking $w(\vec x)$ as the Gaussian 
ground state in the local oscillator potential around any of the sites.
More precisely, $w(\vec x)$ is the exact Wannier wave function of the lowest
band. In a separable periodic potential like that in (6), $w(\vec x)$ decays 
exponentially in all directions rather than in a Gaussian manner [17],
however this does not seriously affect the calculation of $U$ in the 
deep lattice limit $V_0\gg E_r$. From eqn.(10), it is obvious that the 
strength of the repulsion increases with $V_0$ due to the tighter    
squeezing of the on-site wave function $w(\vec x)$. With increasing 
$V_0/E_r$, therefore, not only does the kinetic energy drop
exponentially, but at the same time the interaction energy increases. 
As a result, it is possible to reach the strong coupling regime
$U\approx J$ simply by increasing the depth of the optical lattice 
potential. Regarding the requirements necessary for the validity of
the discrete lattice model (9), it is obvious that the atoms have to 
remain in the lowest vibrational state at each site even in the presence 
of strong interactions. We thus require that $U\ll\hbar\omega_0$ which
is well obeyed even in deep optical lattices as long as $ka_s\ll 1$. 

The zero temperature phase diagram of the homogeneous Bose-Hubbard model
was first discused by Fisher et.al. [16]. Although there are considerable 
quantitative differences between the case of one-, two- or 
three-dimensional lattices, the qualitative structure is similar in all 
cases and is shown schematically in Figure 1. At large $J/U$ the kinetic
energy dominates and the ground state is a delocalized superfluid (SF). 
At small values of $J/U$, interactions dominate and one obtains a series of
so called Mott-insulating (MI) phases with fixed integer filling
$\bar n=1,2,\ldots$ depending on the value of the chemical potential $\mu$.
To understand the peculiar structure of these 'Mott-lobes', consider first
the case of unit filling, i.e. the number $N$ of atoms is precisely 
equal to the number $M$ of 
lattice sites. In the limit where $V_0$ is very large compared to $E_r$, there is
no hopping ($J=0$) and the obvious ground state
\begin{figure}[htb]
\centering\includegraphics[scale=0.6]{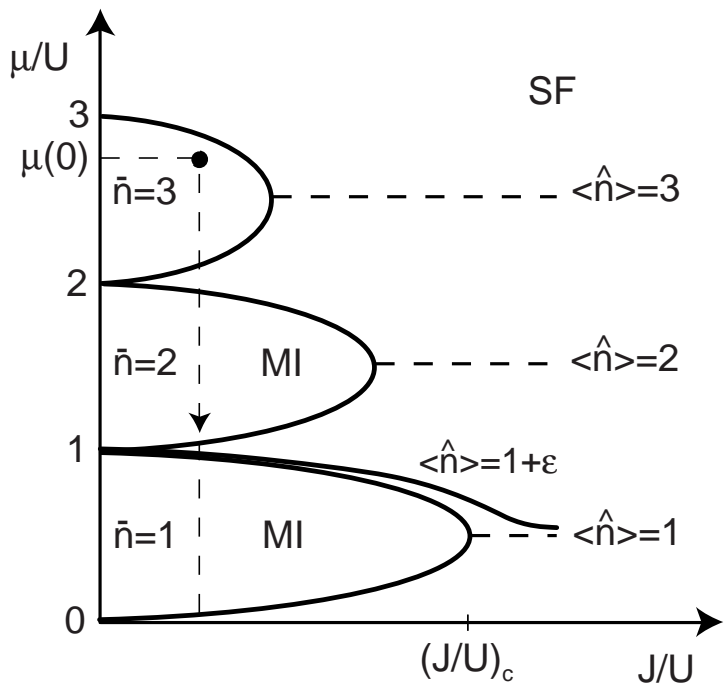}
\caption{Schematic zero temperature phase diagram of the Bose-Hubbard model.
The dashed lines of constant integer density $<\hat n>=1,2,3$ in the superfluid
phase (SF) hit the corresponding Mott-insulating (MI) phases at the tips of the
lobes at a critical value of $J/U$, which decreases with density $\bar n$. For
$<\hat n>=1+\varepsilon$ the line of constant density stays outside the $\bar n
=1$ MI because a fraction $\varepsilon$ of the particles remains superfluid down
to the lowest values of $J$. In an external trap with a $\bar n=3$ MI phase in the 
center,  a series of MI and SF regions appear by going towards the edge of
the cloud, where the local chemical potential has dropped to zero.}
\end{figure}
$$\vert\Psi_{MI} >(J=0, \bar n)=
\prod_{l}\left(\vert\bar n>_{l}\right)\eqno(11)$$     

is a simple product of local Fock-states with precisely one atom ($\bar 
n=1$)
per site. Upon lowering $V_0$, the atoms start to hop around, which 
necessarily involves double occupancy, increasing the energy by $U$. Now as 
long as the gain $J$ in kinetic energy due to hopping is smaller than $U$,
the atoms remain localized although the ground state is no longer a simple
product state as in (11). Once $J$ becomes of order or larger than $U$,
the gain in kinetic energy outweighs the repulsion due to double  
occupancies and the atoms will be delocalized over the whole lattice. In the
limit $J\gg U$ the many body ground state becomes simply an ideal 
Bose-Einstein-condensate where all $N$ atoms are in the $\vec q=0$ 
Bloch-state of the lowest band. Including the normalization factor in a 
lattice with a total number of $M$ sites, this state can be written in the 
form

$$\vert\Psi_{SF,N} >(U=0)=
\left(\frac{1}{\sqrt{M}}\sum_{l=1}^M\, \hat b_l^{\dagger}\right)^N
\,\vert 0>\, .\eqno(12)$$     
 
For large enough $J$ therefore, we recover a Gross-Pitaevski like
description in terms of one, macroscopically occupied state. In two- 
and three-dimensional lattices, the critical
value for the transition from a MI to a SF is reasonably well described  
by a mean-field approximation, giving $(U/J)_c=5.8z$ for $\bar n=1$ and
$(U/J)_c=4\bar nz$ for $\bar n\gg 1$. Here $z$ is the number of nearest
neighbours. In one dimension there are strong deviations from a mean-field
approximation and the corresponding values are $(U/J)_c=3.84$ for 
$\bar n=1$ [18,19] and $(U/J)_c=2.2\bar n$ for $\bar n\gg 1$. The
latter result follows from mapping the Bose-Hubbard model to a chain of 
Josephson junctions, for which the critical value of the transition to
a MI phase is known precisely.  
 
Consider now a filling with $<\hat n>=1+\varepsilon$ which is slightly larger
than one. For large $J/U$ the ground state has all the atoms delocalized 
over the whole lattice and the situation is hardly different from the 
case of unit filling. Upon lowering $J/U$, however, the line of constant 
density remains slightly above the $\bar n=1$ 'Mott-lobe',
and stays in the SF regime down to the lowest $J/U$ (see Fig.1). For any 
noninteger filling, therefore, the ground state remains SF as long as the 
atoms can hop at all. This is a consequence of the fact, that even for
$J\ll U$ there is a small fraction $\varepsilon$ of atoms which remain SF
on top of a frozen MI-phase with $\bar n=1$. Indeed this fraction
can still gain kinetic energy by delocalizing over the whole lattice
without being blocked by the repulsive interaction $U$ because two
of those particles will never be at the same place. The same argument applies to
holes when $\varepsilon$ is negative. 

In order to describe the situation in a weak harmonic trap, we use the 
standard appproximation that a slowly varying external potential may be 
accounted for by a spatially varying chemical potential
$\mu_l=\mu(0)-\epsilon_l$ (we choose $\epsilon_l=0$ at the trap 
center). Assuming that the chemical potential $\mu(0)$ at trap center
falls into the $\bar n=3$ 'Mott-lobe', one obtains a series of 
MI domains separated by a SF by moving to the boundary of the trap
where $\mu_l$ vanishes (see Fig.1) [15]. In this manner, all the different
phases which exist for given $J/U$ below $\mu(0)$ are present 
simultaneously. Since the defining property of a MI-phase is its
incompressibility $\partial n/\partial\mu =0$, the atomic density stays
constant in the Mott-phases, even though the external trapping potential is 
rising. An estimate for the width of the incompressible domains is
obtained by noting that for $J\ll U$ the range in chemical potential 
over which the density remains constant is close to $U$. In a quadratic 
confining potential with axial frequency $\nu_z\approx 40$Hz and
with typical values $U/h=1$kHz, the width of the incompressible 
MI-states is around $10\mu$m. It remains an experimental  challenge to
spatially resolve the SF and MI phases in a trap, thus verifying the 
crucial property of incompressibility. 

In practice, the observation of the SF to MI transition is done in the 
usual manner by absorption imaging the atomic cloud after a given
expansion time. The corresponding series of images is shown in Fig.2 
for different values of $V_0$, ranging between $V_0=0$ (a) and 
$V_0=20E_r$ (h). One observes a series of 'Bragg-peaks' around the
characteristic 'zero-momentum' peak of a condensate in the absence
of an optical lattice. With increasing $V_0$ these peaks become more 
pronounced. Beyond a critical lattice depth around $V_0=13E_r$ (e), this 
trend is suddenly reversed, however, and the 'Bragg peaks' eventually 
disappear completely. In order to understand, to which extent these 
pictures actually provide a direct evidence for the existence of a SF
to MI transition predicted by the Bose-Hubbard model, we neglect
the inhomogeneous nature of the atomic cloud and assume that the 
absorption images simply reflect the momentum distribution. For atoms
which are confined to move in the lowest band of the lattice, it is 
straightforward to show that the momentum (not quasi-momentum [20]) 
distribution 
\begin{figure}[htb]
\centering\includegraphics[scale=1.2]{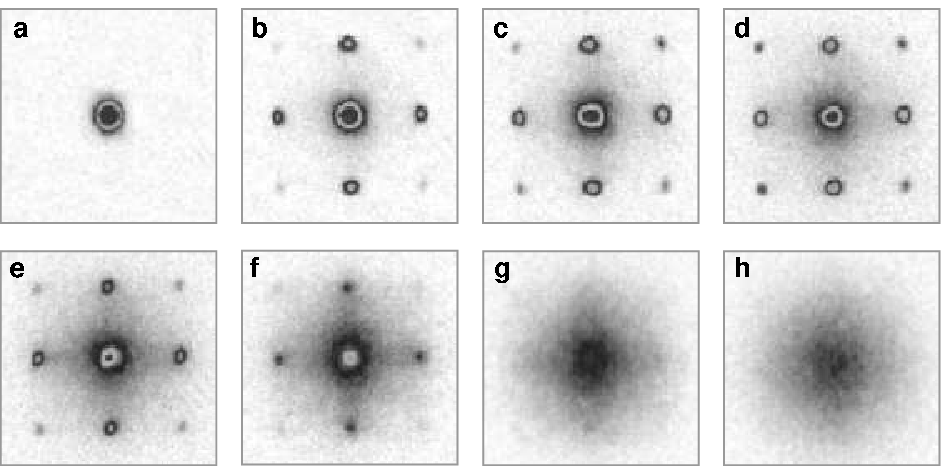}
\caption{Absorption images of the interference pattern from atoms in an
optical lattice  with varying values of the potential depth $V_0/E_r=$ 
$0$ (a), $3$ (b),  $7$ (c), $10$ (d), $13$ (e), $14$ (f), $16$ (g) and
$20$ (h). Taken from ref. [14] with permission.}
\end{figure}
$$n(\vec k)=n\vert w(\vec k)\vert^2\,\sum_{\vec R}e^{i\vec k\cdot\vec R}
\rho_1(\vec R)\eqno(13)$$

can be expressed in terms of the exact one-particle density matrix
$\rho_1(\vec R)=<\hat b_{\vec R}^{\dagger}\hat b_{0}>$ at separation
$\vec R$  and the Fourier transform $w(\vec k)$ of the associated
Wannier wave-function. The summation in (13) is over all lattice
vectors $\vec R$, which are integer multiples of the three primitive
vectors of the given lattice. Now the SF and MI phases are distinguished
quite generally by the behaviour of the one particle density matrix 
(or first order coherence function in quantum optics terminology) 
at large separation.
In the SF, $\rho_1(\vec R)$ approaches a finite value
$\lim_{|\vec R|\to\infty}<\hat b_{\vec R}^{\dagger}\hat b_{0}>
=n_0/n$ which defines the condensate density $n_0$ [21]. For the MI phase,
in turn, $\rho_1(\vec R)$ decays to zero exponentially. Using (13), the 
SF phase of cold atoms in an optical lattice can thus quite generally be 
characterized by the fact that at reciprocal lattice vectors $\vec k=\vec 
G$ defined by $\vec G\cdot\vec R=2\pi$ times an integer, the
momentum distribution $n(\vec k=\vec G)$ has a peak 

$$n(\vec k=\vec G)=N\cdot n_{0}\vert w(\vec G)\vert^2
\eqno(14)$$

which scales with the
total number $N$ of particles. This is the expected behaviour for the 
interference pattern from a periodic array of phase coherent sources of 
matter waves and is precisely analogous to the more standard Bragg-peaks
in the static structure factor of a solid, with the condensate fraction
playing the role of the Debye-Waller factor [22]. The fact that the
peaks in the momentum distribution at $\vec k=\vec G$ initially grow
with increasing depth of the lattice potential is a result of the 
strong decrease in spatial extent of the Wannier function $w(\vec x)$, which
entails a corresponding increase in its Fourier transform $w(\vec k)$
at higher momenta. It is important to realize that there is no broadening 
of the peaks as long as $\rho_1(|\vec R|\to\infty)$ is finite, in agreement
with the experimental observations [14]. 
In the MI regime, where $\rho_1(\vec R)$ decays to
zero, remnants of the 'Bragg-peaks' still remain (see e.g. (f) in Fig.2)
as long as $\rho_1(\vec R)$ extends over several lattice spacings, because 
the series in (13) adds up constructively at $\vec k=\vec G$. Physically
this reflects the fact that phase coherence is still present over
distances much larger than one lattice spacing provided one is close
to the transition to superfluidity. In contrast to the SF regime, however, these
peaks are now broadened and do not scale with the total number $N$ of particles. 
In the extreme MI limit $J\ll U$, 
hopping of atoms completely vanishes and $\rho_1(\vec R)$ is zero beyond
$\vec R=0$. Coherence is then completely lost and the momentum distribution
is a structureless Gaussian, reflecting the Fourier transform of the
Wannier wave function (see (h) in Fig.2). These arguments show that for 
a large and homogeneous system there is indeed a sharp signature of the
SF to MI transition in the interference pattern. It is connected
with the existence or not of (off-diagonal) long range order in the 
one particle density matrix, which effectively measures the range of
phase coherence and the condensate fraction. Of course the actual system is 
not homogeneous and a numerical computation of the interference pattern is 
necessary for a quantitative comparison with experiment [23]. Due to the
finite size and the fact that different MI phases are involved, the pattern 
evolves continously from the SF to the MI regime. Indeed, as is evident 
from the phase diagram in Fig.1, the critical value of $J/U$ is different 
for the two different MI phases $\bar n=1$ and $\bar n=2$ which are present 
in the trap. Nevertheless, a rather sharp transition is observed  
experimentally, because $J/U$ depends exponentially on the control
parameter $V_0/E_r$. The small change from $V_0=13E_r$ in (e) to 
$V_0=14E_r$ in (f) thus covers a range in $J/U$ wider than that which would 
be required to distinguish the $\bar n=1$ from the $\bar n=2$ transition.

A second signature of the SF to MI transition is the appearance of a 
finite
excitation gap $\Delta\not= 0$ in the MI. Deep in the MI phase, this gap
has size $U$, which is just the increase in energy if an atom tunnels to
an already occupied adjacent site (note that this is much smaller than
the gap $\hbar\omega_0$ for the excitation of the next vibrational state). 
The existence of this gap has been 
verified experimentally by applying a phase gradient in the MI and 
measuring the resulting excitations produced in the SF at smaller
$V_0/E_r$ [14]. In this manner the fact that $\Delta(J\ll U)=U$ was
verified for a range of $V_0$, all reasonably deep in the MI phase. 
For reasons discussed above, however, it has not been possible to see
the vanishing of the gap near the transition, which should  scale like
$\Delta\sim(J_c-J)^{1/2}$ in the three-dimensional case [16]. In the SF
regime, there is no excitation gap and instead the homogeneous system
exhibits a sound like mode with frequency $\omega(q)=cq$. The associated
sound velocity follows from the thermodynamic relation $mc^2=n_s\cdot
\partial\mu/\partial n$ and thus gives information about the superfluid
density $n_s$. The existence of a sound like excitation even in the 
presence of an underlying lattice which explicitely breaks translation
invariance is a consequence of long range phase coherence in the SF. Its
observation would thus constitute an independent proof that the atoms 
move coherently over the whole lattice and thus phase gradients give rise
to dissipationless currents.

Finally we discuss the change in the atom number statistics at individual
sites between the SF and the MI regime. This issue has been investigated
in a very recent beautiful experiment, observing collapse and revival
of the matter wave due to the coherent superposition of states with 
different atom numbers in the SF [24]. As noted above, the ground state 
(11) in the extreme MI limit is a product of Fock states with a definite 
number $\bar n$ of particles at each site. At finite hopping $J\not= 0$,
this simple picture breaks down because the atoms have a finite amplitude
to be at different sites. The many body ground state can then no longer
be written as a simple product state as in (11). In the opposite limit
$U\to 0$, the ground state is a condensate of zero
quasimomentum Bloch states. It turns out, that the probability
of finding precisely $n$ atoms at any given site in the associated state 
(12) is close to a Poissonian distribution. 
More precisely, in the limit $N,M\to\infty$ at fixed 'density' $N/M$,
the state (12) becomes indistinguishable in a local measurement from
a coherent state

$$\vert\Psi_{SF}>(U=0)=\exp{\sqrt{N}\hat b_{q=0}^{\dagger}}\vert 0>=
\prod_{l=1}^M \left(\exp{\sqrt{\frac{N}{M}}\hat b_l^{\dagger}}\vert 0>_l
\right)\eqno(15)$$

which factorizes into a product of local Poissonian states with average
$<\hat n_l>=N/M$ because boson operators at different sites commute. We
have thus come to the remarkable conclusion that for integer densities
$N/M=\bar n=1,2,\ldots$ the many body ground state may be written in a 
local product form

$$\vert\Psi_{GW} >\, =
\prod_{l}\left(\sum_{n=0}^{\infty}c_{n}\vert n>_{l}\right)\eqno(16)$$

in both limits $J\to 0$ and $U\to 0$. The associated atom number 
probability distribution $p_n=|c_n|^2$ is either a pure Fock or a full
Poissonian distribution. It is now very plausible to use the factorized
form (16) as an approximation for arbitrary $J/U$, taking the coefficients
$c_n$ as variational parameters which are determined by minimizing the 
ground state energy [25]. As first pointed out by Rokhsar and Kotliar [26],
this is effectively a Gutzwiller ansatz for bosons. Beyond being 
very simple computationally, this ansatz describes the SF to MI 
transition in a mean-field sense, becoming exact in infinite
dimensions. In addition, it provides 
one with a very intuitive picture of the transition to a MI state,
which occurs precisely at the point where the local number distribution
becomes a pure Fock distribution. This is consistent with a vanishing
expectation value of the local matter wave field

$$<\Psi_{GW}\vert\hat b_l\vert\Psi_{GW}>=\sum_{n=1}^{\infty}\,
\sqrt{n}c_{n-1}^*c_n\eqno(17)$$

in the Gutzwiller approximation. It is important, however, to emphasize 
that the ansatz (16) fails to account for the nontrivial correlations
between different sites present at any finite $J$. These correlations
imply that the one particle density matrix $\rho_1(\vec R)$ is 
different from zero at finite distance $|\vec R|\not= 0$,
becoming long ranged at the transition to a SF. By contrast, in the 
Gutzwiller approximation, the one particle density matrix has no
spatial dependence at all: it is zero at any $|\vec R|\not= 0$ in the 
MI and is completely independent of $\vec R$ in the SF. Moreover, in the
Gutzwiller approximation the phase transition is directly reflected in the
local number fluctuations, with the variance of $n_l$ vanishing
throughout the MI phase. By contrast, in an exact theory local variables
like the on-site number distribution will change in a smooth manner near the
transition and the variance of the local particle number will only
vanish in the limit $J\to 0$. Concerning the dynamics, one expects 
that the Gutzwiller approximation qualitatively captures the time scales
for local changes of the configuration, however it fails to correctly 
describe long wavelength excitations [26].

The realization of a SF to MI transition with cold atoms
in optical lattices provides an essentially perfect realization of one of
the most prominent models in many body physics. It allows to study
a quantum phase transition by simply tuning the depth of the optical 
lattice. There remain, however, a number of open questions
in particular concerning the detailed spatial
structure in the trap and the dynamical behaviour [27].

\section{Crystallization in weak optical lattices}

As we have discussed above, superfluidity of cold atoms is destroyed
in a deep optical lattice where the ground state looses phase
coherence and - essentially - has a fixed number of atoms per site.
The associated Mott-Hubbard transition occurs when the ratio
$U/J$ is of order one. Using the expressions (8) and (10)
for $J$ and $U$ in terms of the optical lattice parameters, this
translates into a condition of the form

$$\frac{a_s}{a}\cdot\exp{2\left(V_0/E_r\right)^{1/2}}=
\frac{\sqrt{2}}{\pi}\cdot\left( U/J\right)_{c}\eqno(18)$$

for the critical value of the dimensionless lattice depth $V_0/E_r
\vert_c$. In the interesting regime with one or two atoms per site, the
lattice constant $a$ is roughly equal to the the mean interparticle spacing
$n^{-1/3}$. The prefactor $a_s/a$ in (18) thus coincides with the 
dimensionless interaction parameter $\gamma$ introduced in (2). For 
weak interactions $\gamma\ll 1$ therefore, the SF to MI transition 
requires deep optical lattices, for instance $V_0\vert_c\approx
13E_r$ in the experiments by Greiner et.al.[14], where $a_s/a\approx 0.01$.
In the following we want to adress the question what happens in a 
situation where the effective gas parameter $\gamma$ becomes of order
one or larger. The superfluid ground state is then expected to be
destroyed already in a weak optical lattice, where the description in terms 
of a Bose-Hubbard model is no longer applicable. It turns out that this 
problem can be solved completely in the special case of one-dimensional
Bose gases in which the transverse motion is frozen into the lowest eigenstate
of a strong confining potential [28].  As pointed out by Petrov et.al.
[29], the ratio between the interaction and kinetic energy per particle 
in one dimension

$$\gamma_1=\frac{g_1n_1}{\hbar^2n_1^2/m}=\frac{2a_s}{n_1l_{\perp}^2}
\eqno(19)$$

scales inversely with the 1d density $n_1$. Here $g_1$ is the strength
of the effective delta-function interaction in 1d and $l_{\perp}$ the
oscillator length for the transverse confinement [29]. In one dimension,
it is thus the low density limit where interactions dominate. This
somewhat counterintuitive result can be understood physically by noting that
at low 1d densities, the average kinetic energy per particle
$\epsilon_{kin}\sim n_1^2$ vanishes so quickly that the atoms are 
perfectly reflected by the repulsive potential of the surrounding 
particles. For $\gamma_{1}\gg 1$, therefore, the system aproaches a gas
of impenetrable bosons which is called the Tonks-limit [29]. In particular
at $\gamma_{1}=\infty$ the exact many body wave function is just the absolute 
value of that of a free Fermi gas, as was shown a long time ago by 
Girardeau [30]. This equivalence remains valid in the presence of an
arbitrary additional one-particle potential like that of an optical
lattice. For a qualitative understanding of what happens in the strongly
interacting regime it is therefore useful to consider a free Fermi gas
in a weak periodic potential $V(x)=V_0\sin^2{kx}$ with $V_0$ of order $E_r$ 
or smaller. This is an elementary problem in solid state 
physics, equivalent to the nearly free electron limit of a one-dimensional
bandstructure. The single particle spectrum consists of a series of free 
particle like
bands separated by energy gaps $\Delta_l \; l=1,2\ldots$. The gaps become 
exponentially 
small with increasing energy, scaling like $\Delta_l\sim\vert V_0\vert^l$ 
in the limit
$V_0\ll E_r$.  For a commensurate density, where an integer number $l$ 
of particles
fit into one unit cell, the $l$ lowest bands are completely filled. 
The groundstate of
noninteracting fermions is thus a trivial band insulator.  Similar to the 
incompressible
Mott-insulating phase of the Bose-Hubbard model, the state with a fixed integer 
density remains locked over a finite range $\Delta_l$ of the chemical potential.
For weak optical lattices $V_0\ll E_r$ the lowest gap 
$\Delta_{l=1}=\vert V_0\vert/2$
is much larger than the higher order ones.  As  a result, it is the 
commensurate phase with unit filling $N=M$ which has maximal stability.

In the Tonks limit $\gamma_1\to\infty$ we have thus found that an arbitrary 
weak optical 
lattice which is commensurate with the average density will pin the atoms into 
an incompressible optical crystal. The crucial question is obviously whether this 
peculiar feature of hard core bosons in one dimension is still present at 
finite and
experimentally realizable values of $\gamma_1$. To answer that, it is 
convenient to
use Haldane's description of Bose gases with arbitrary repulsive, short range
interactions in terms of their long wavelength density oscillations [31].  
Introducing
a field $\phi(x)$ which is related to the fluctuations $\delta n(x)$ 
around the average 
density via $\delta n(x)=\beta\,\partial_x\phi/2\pi$, the Hamiltonian 
in the presence of 
a weak, commensurate optical lattice can be shown [28] to be that of a quantum
$(1+1)-$dimensional sine-Gordon model      

$$\hat H=\frac{\hbar v_{s}}{2}\int dx 
\left[\hat\Pi^2(x)+(\partial_{x}\hat\phi)^2+\frac{2V_{0}n_1}{\hbar v_{s}}
\cos{\beta\hat\phi}\right]\eqno(20)$$

Here $v_s$ is the actual sound velocity and $\hat\Pi(x)$ is canonically conjugate
to $\hat\phi(x)$ such that $\left[\hat\phi(x),\hat\Pi(x')\right]=i\delta(x-x')$. 
The coupling
parameter $\beta$ is related to the dimensionless ratio $K=\pi\hbar n_1/mv_s
=\beta^2/4\pi$ which characterizes the power in the characteristic decay 

$$\rho_1(x, T)\sim\left(\frac{\hbar v_s}{k_BT}
\sinh{\frac{\pi\vert x\vert k_BT}{\hbar v_s}}\right)
^{-1/2K}\eqno(21)$$

of the one particle density matrix in the absence of the optical lattice, 
typical for one-dimensional quantum liquids [31].
In Haldane's description, the correlation exponent $K$ is a phenomenological
parameter which approaches $K\to 1$ for hard core Bosons and $K\to\infty$ in
the ideal gas limit. For given $K,n_1$ and strength $V_0$ of the optical lattice,
the sine-Gordon model (20) is an exactly soluble field theory [32]. It exhibits a 
transition at a critical value $\beta_c^2=8\pi\; (K_c=2)$ such that for $K>2$ the
Bose gas ground state remains gapless and superfluid in a weak optical
lattice while for $(1<)K<2$ the atoms are locked even in an arbitrary weak 
periodic lattice as long as the deviation $Q=2\pi (n_1-1/a)$ between the 
period $a$ and
the average interparticle spacing $n_1^{-1}$ is less than a critical value $Q_c$.
The commensurate phase is characterized by a finite excitation gap [28]

$$\Delta=\frac{2E_r}{K}\left(\frac{K\vert V_0\vert}{(2-K)4E_r}\right)^{1/(2-K)}\eqno(22)$$ 
 
which is nonanalytic in $V_0$. It approaches $\vert V_0\vert /2$ in the Tonks 
gas limit $K\to 1$ in agreement with the ideal Fermi gas picture 
discussed above. The size of the gap also determines the critical
value $Q_c=K^2\pi n_1\Delta/2E_r$ of the deviation from exact commensurability
which can still be accomodated into a locked groundstate. 
In order to relate $K$ to the microscopic and 
experimentally tunable parameter
$\gamma_1$ introduced in (19), we use the exact solution by Lieb and Liniger
[33] of the 1d Bose gas with a delta-function interaction of strength $g_1$.
It turns out that $K(\gamma_1)$ is a monotonically decreasing function
which reaches the critical value $K_c=2$ at $\gamma_{1,c}\approx 3.5$.
The transition to a commensurate, incompressible state in a weak
optical lattice thus occurs long before the Tonks limit is reached. As is evident
from eqn. (22), however, the gap is exponentially small in the vicinity of
the critical value $K_c=2$.  In order to reach appreciable values of the gap,
one thus needs $\gamma\approx 10$ where $K\approx 1.4$.
   
In a one-dimensional Bose gas we have thus found that a transition from a SF to 
a MI state at weak coupling $\gamma\ll 1$ requires a 
deep optical lattice, while at strong coupling $\gamma\gg 1$ an 
arbitrary weak lattice is sufficient to destroy phase coherence
(similar to the MI phase of the Bose-Hubbard model, the one particle 
density matrix
$\rho_1(x)$ will decay exponentially in the phase where the atoms are locked
to the external lattice potential). For weak coupling,  the criterion 
$U/J\vert_c=2C\approx 3.84$ for the transition in the anisotropic,
one-dimensional 
Bose-Hubbard model can be written in a  form similar to eqn. (18)

$$\frac{4V_0}{E_r}=\ln^2\left[4\sqrt{2}\pi C(V_0/E_r)^{1/2}/\gamma\right] . 
\eqno(23)$$

The solution of this transcendental equation gives a critical value 
$V_0/E_r\vert_c$ which increases rather slowly as a function of
the inverse interaction parameter $1/\gamma$.  From the exact solution
of the sine-Gordon model, in turn, we know that  - at least in one dimension -
this critical amplitude of the optical lattice vanishes at the finite value
$1/\gamma\vert_c\approx 0.29$.  More precisely, the 
Kosterlitz-Thouless nature of the transition near $K_c=2$ determines
the critical value of $K$ for small $u=KV_0/4E_r$ to behave like $K_c(u)=2(1+u)$
to linear order in $u$ [28].   The complete phase diagram for unit filling 
at arbitrary
values of $V_0/E_r$ can then be obtained by combining these two 
asymptotic results in a smooth interpolation, as shown in Fig.3.  The 
Bose-Hubbard transition (BH) for weakly interacting gases in a deep
optical lattice thus continuously evolves into one of the 
commensurate-incommensurate
(C-IC) type for strongly interacting gases in a weak lattice.  It remains 
a challenge
to extend these results to situations with two or more particles per lattice 
period and - in particular - to the case of two- or three-dimensional lattices.

\begin{figure}[htb]
\centering\includegraphics[scale=0.6]{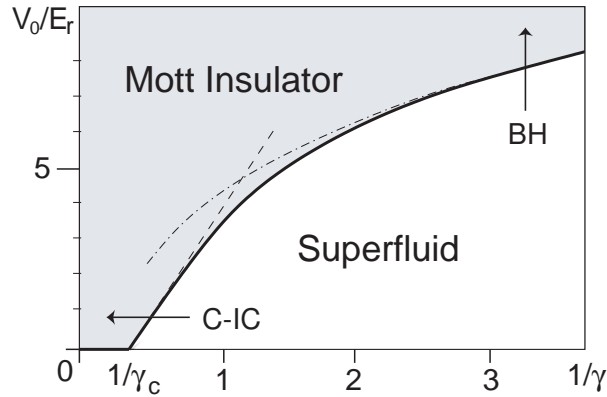}
\caption{Ground state  phase diagram of a one-dimensional Bose gas in 
an optical lattice of arbitrary strength $V_0/E_r$ as a function of the
inverse interaction parameter $\gamma$. The dashed line is the asymptotic
behaviour near the critical point $\gamma_c$ as determined from the 
sine-Gordon model, while the dashed-dotted line derives from the
solution of the transcendental equation (23) as obtained from
the critical point of the 1d Bose-Hubbard model.}
\end{figure}

Regarding the prospects for an experimental observation of the C-IC transition
discussed above, we have seen that realistic values of the excitation gap
$\Delta$ require $\gamma_1$ to be of order $10$. One-dimensional Bose gases
with parameters in this range may be realized with a strong optical lattice
in only two directions $x,y$ which confine the atoms transversely but leave 
the motion along $z$  essentially free, except for a rather weak axial trap with
frequency $\nu_z$. As an example, using numbers close to those in the 
experiments by Greiner et.al. [14] on the SF to MI transition, it is possible
to generate a few thousand parallel one-dimensional gases with about 
$N=50$ atoms per wire.  Taking realistic values $\nu_{\perp}=20$ kHz
and $\nu_z=40$ Hz, the central density for large $\gamma$ is close to 
$n_1(0)=2\mu$m$^{-1}$ [34]. This is precisely commensurate with the lattice
constant $a\approx 0.5\mu$m of the  optical lattice used in these experiments. 
Adding a weak optical lattice in $z-$direction, will thus lead to an 
incompressible 
state in the center of the trap provided $\gamma_1$ is larger than the critical
value $\gamma_{1,c}\approx 3.5$.  For $^{87}$Rb with a scattering length
$a_s\approx 5$nm, the resulting $\gamma_1$ is close to one and thus not
in the required range. With a tunable scattering length as in $^{85}$Rb, however,
it is perfectly feasible to increase $a_s$ by one order of magnitude and thus
a Mott-insulating state could be realized with a very weak optical lattice.
Since three-body losses are very strongly reduced in one dimension at
large $\gamma$ [35], there is no problem with the condensate lifetime 
here.  
The transition may be observed in a similar manner than in the Bose-Hubbard
case. More interesting however would be to measure the long range
translational order present in the locked phase by Bragg diffraction, as 
was done for cold atoms in deep optical lattices even at very low
densities [22].  This method has the advantage that the signal from many
parallel wires adds up constructively because they all experience the
same modulation in $z-$direction.

\section{Conclusion and Outlook}

With the recent realization of a quantum phase transition between a superfluid
and a Mott-insulating state by Greiner et.al. [14], the field of cold atoms has
entered a regime, where strong correlation effects may be studied in an
unprecedentedly clean manner. Indeed, basic models in many body theory
like the Hubbard-model for bosons or fermions with on-site interaction,
which were originally introduced in a condensed matter context as a rather
schematic description of say superfluid Helium in Vycor or electrons
in high-temperature superconductors can now be applied even on
a quantitative level. Moreover the crucial parameters $J,U$ and density can
easily be tuned in a controlled fashion. This opens a wide area of possibilities
for strong correlation physics with cold atoms, in particular if degenerate 
fermions may be loaded into an optical lattice. With two equivalent species
of fermions the resulting version of the Hubbard model displays a wealth
of different phases: in the attractive case it describes the BCS- to Bose-crossover
for Cooper-pairing, in the repulsive case antiferromagnetic or unconventional
superconducting phases appear, as recently discussed by Hofstetter et.al. [36].
From the perspective of atomic and molecular physics, a Mott-insulating state
with precisely two atoms per site is an ideal starting point for the formation of
molecular condensates via local photoassociation and subsequent melting
of the Mott-phase  as suggested by Jaksch et.al. [37]. Finally, controlled
interactions ('collisions')  between atoms in different internal states in an
optical lattice may be used to generate highly entangled states useful for
quantum computation schemes [38,39]. Cold atoms in optical lattices have 
therefore started to fascinate people not only in the
field of atomic physics and quantum optics but far beyond that and it seems
that the field is at the beginning of a promising area in 
research.

\ack{
It is a pleasure to acknowledge many discussions with I. Bloch, M. Greiner, 
T. Esslinger and M. Weidem\"uller on the experimental side and with
P. Fedichev, A. Recati and P. Zoller on theory.  The work on crystallization 
in weak 
optical lattices was done together with H.P. B\"uchler and G. Blatter during
a sabbatical at the ETH in Z\"urich, whose great hospitality is gratefully
acknowledged.}

\end{document}